\documentclass[pre,twocolumn,a4paper,showpacs,superscriptaddress]{revtex4}

\usepackage{graphicx,amsmath,times}
\usepackage[english]{babel}

\begin{document}

\title{The effect of confinement on the deformation of microfluidic drops}
\author{Camilo Ulloa}
\affiliation{Departamento de F\'isica, Facultad de Ciencias F\'isicas y Matem\'aticas, Universidad de Chile, Av. Blanco Encalada 2008, Santiago, Chile}
\author{Alberto Ahumada}
\affiliation{Universit\'e Paris-Est Marne-La-Vall\'ee, 5 boulevard Descartes, 77545 Marne-La-Vall\'ee Cedex 5, France}
\author{Mar\'ia Luisa Cordero}
\email{mcordero@ing.uchile.cl}
\affiliation{Departamento de F\'isica, Facultad de Ciencias F\'isicas y Matem\'aticas, Universidad de Chile, Av. Blanco Encalada 2008, Santiago, Chile}

\date{\today}

%%%%%%%%%%%%%%%%
% Abstract
%%%%%%%%%%%%%%%%

\begin{abstract}
We study the deformation of drops squeezed between the floor and ceiling of a microchannel and subjected to a hyperbolic flow. We observe that the maximum deformation of drops depends on both the drop size and the rate of strain of the external flow and can be described with power laws with exponents $2.59 \pm 0.28$ and $0.91 \pm 0.05$ respectively. We develop a theoretical model to describe the deformation of squeezed drops based on the Darcy approximation for shallow geometries and the use of complex potentials. The model describes the steady-state deformation of the drops as a function of a non-dimensional parameter $Ca \, \delta^2$, where $Ca$ is the capillary number (proportional to the strain rate and the drop size) and $\delta$ is a confinement parameter equal to the drop size divided by the channel height. For small deformations, the theoretical model predicts a linear relationship between the deformation of drops and this parameter, in good agreement with the experimental observations.
\end{abstract}


\maketitle

%%%%%%%%%%%%%%%%
% Introduction
%%%%%%%%%%%%%%%%

\section{\label{sec:Introduction}Introduction}

Drops are usually confined to flow in shallow geometries and this is often the case in droplet-based microfluidics~\cite{Dollet2008, Hashimoto2008}. Inside microchannels, drops adopt a discoidal shape of characteristic radius $R_o$ and height $h$, which determines the degree of confinement through the parameter $\delta = 2R_o/h > 1$. It is to expect that confinement plays a key role in the dynamics of drops. Moreover, this geometrical constraint can be useful for the theoretical description of microfluidic drops since the flow of fluid at low Reynolds number in Hele-Shaw cells can be simplified into a two-dimensional problem~\cite{Batchelor, Park1984}. However, little is known about the difference that confinement introduces in the behavior of drops versus the cases without confinement. For example, it is known that the deformation of unconfined drops in shear flows can be described in terms of two non-dimensional parameters, the capillary number $Ca$ and the viscosity ratio $\lambda$~\cite{Taylor1934, Rallison1984, Stone1994}. When taking into account the confinement, a model describing the deformation of squeezed drops should include all three parameters, $Ca, \lambda$ and $\delta$.

Thus far, the effect of confinement on the deformation and breakup of drops has been studied in cases with $\delta < 1$~\cite{Migler2001, Pathak2003, Sibillo2006}, in which the presence of nearby walls alters the deformation of non-squeezed drops. A few experiments have been conducted to study the deformation of squeezed drops in microchannels~\cite{Cubaud2009, Mulligan2011, Salkin2012}. However, these studies remain very qualitative and a more quantitative description of the role of confinement in the deformation of drops is still lacking. In this article we study the effect of confinement on the drop deformation in cases with $\delta > 1$. We perform controlled experiments where drops are subjected to a hyperbolic flow to measure their deformation under confinement. We propose a theoretical model based on the Darcy approximation for shallow geometries~\cite{Batchelor, Park1984}. Together, these results demonstrate that the effect of confinement can be quantified in a simple analytical way through the confinement parameter $\delta$.

The article is organized as follows. In Sec.~\ref{sec:Experimental} we describe the experimental setup and the experimental results are presented in Sec.~\ref{sec:Results}. The theoretical model that explains the behavior of a squeezed drop in a hyperbolic flow in presented in Sec.~\ref{sec:Theory}. Finally, experimental and theoretical results are compared and discussed in Sec.~\ref{sec:Discussion}.

%%%%%%%%%%%%%%%%
% Methods
%%%%%%%%%%%%%%%%

\section{\label{sec:Experimental}Materials and methods}

\subsection{\label{subsec:Setup}Experimental setup}

Experiments are performed inside microchannels fabricated in polydimethylsiloxane (Sylgard 184, Dow Corning) using standard soft lithography techniques~\cite{McDonald2000}. Microchannels are treated with a siliconizing agent (Sigmacote, Sigma-Aldrich) prior to their use in order to improve their hydrophobicity.

\begin{figure}[h]
\includegraphics[width=8.5cm]{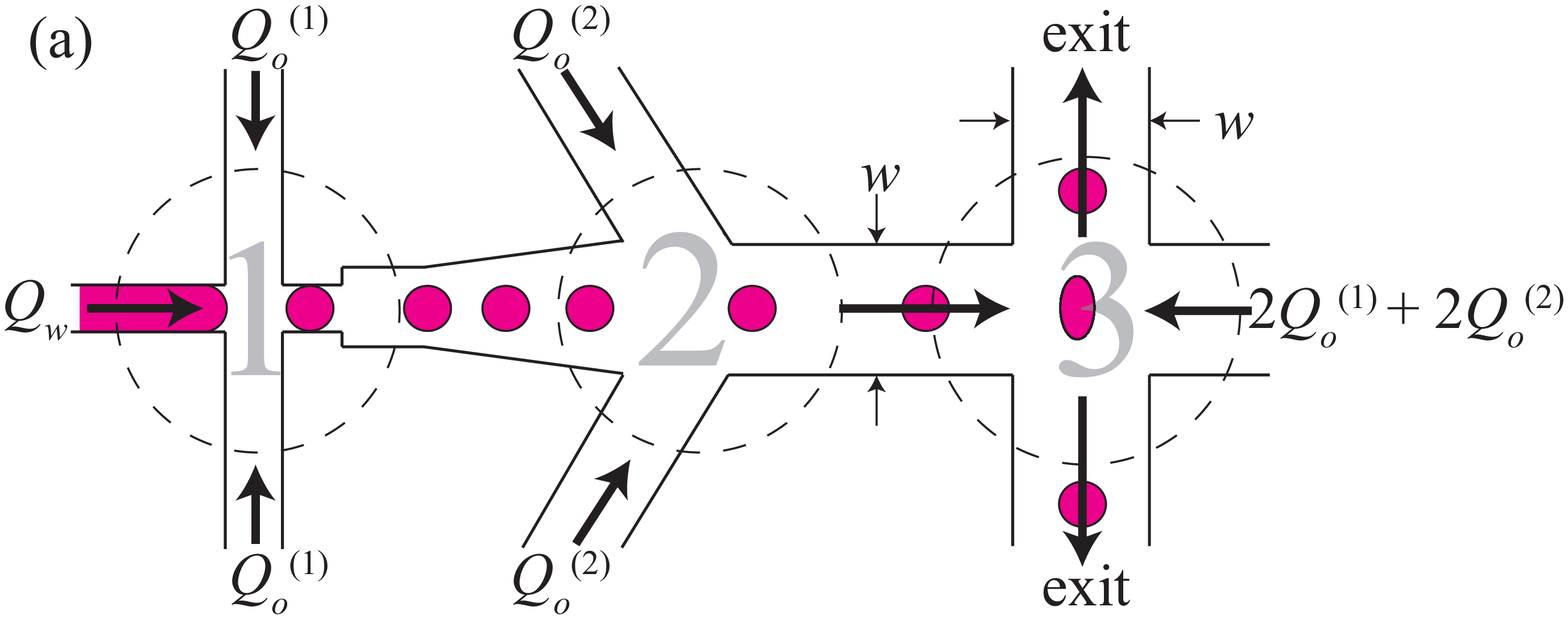}
\includegraphics[width=7cm]{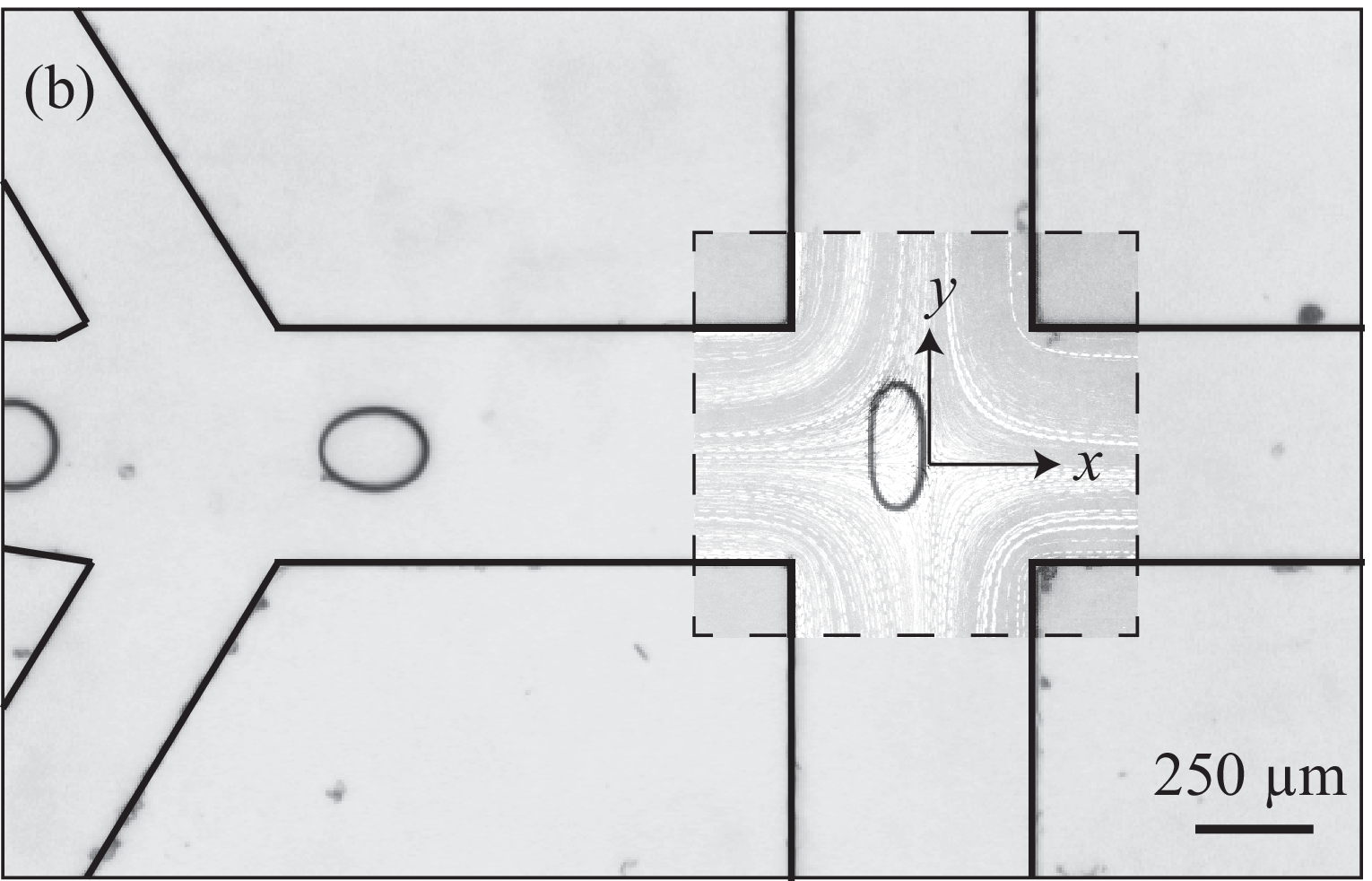}
\caption{\label{fig:Channel}(Color online) (a) Schematic of the microchannel geometry (not to scale), showing the section for drop formation (1), the second oil inlet for the control of the flow velocity (2) and the test section in which drops deform (3). Thick arrows represent the direction of flow. The width at the test section ($w$) is indicated and the height $h$ of the channel is in the direction perpendicular to the figure. (b) Micrography of the channel, with a drop being accelerated by the injection of $Q_o^{(2)}$ and another one being deformed in the test section. The flow in the test section is visualized with fluorescent beads in the absence of drops and several photographs are superimposed to represent the oil streamlines.}
\end{figure}

The channel geometry is schematically presented in Fig.~\ref{fig:Channel}(a). It consists of three modules, the first one for drop formation, the second one for controlling the drop velocity and the third one is the test section where drops deform in response to an extensional flow. The channel geometry is characterized by its height $h$ and its width at the test section $w$. Two channels are used, channel I with $h = 40 \ \mu$m and $w = 300 \ \mu$m and channel II with $h = 58 \ \mu$m and $w = 400 \ \mu$m.

In the first module of the channel, a microfluidic flow focusing~\cite{Anna2003} produces water drops in a carrier oil stream. The produced disk-like drops are characterized by their undeformed radius $R_o$, which is determined by the water and oil flow rates injected in the flow focusing, $Q_w$ and $Q_o^{(1)}$ respectively. Drop radii $R_o$ range between $50 \ \mu$m and $120 \ \mu$m and therefore the confinement parameter $\delta \in [1.7, 6]$.

In the second module, two oil streams (flow rates $Q_o^{(2)}$) are symmetrically injected downstream of the drop formation section to adjust the velocity of drops independently of their size. The total flow rate is defined as $Q_{\rm tot} = Q_w + 2Q_o^{(1)} + 2Q_o^{(2)} \approx 2\left(Q_o^{(1)} + Q_o^{(2)}\right)$ since the water flow rate is much smaller than the oil flow rates.

In the third module, an extensional oil flow is produced to induce drop deformation. For that, the same total oil flow rate $Q_{\rm tot}$ is injected into the channel in the opposite direction and outlet channels are arranged perpendicularly to the main channel. The resulting cross-shaped intersection, where a hyperbolic flow is produced, defines the test section where the deformation of the drops is measured. In the optical micrography of the channel geometry shown in Fig.~\ref{fig:Channel}(b), the hyperbolic flow produced in the test section in the absence of drops is evidenced with fluorescent tracers.

Distilled water is used to produce the drops while the oil phase consists of pure mineral oil (heavy, Sigma-Aldrich) used as received. The water viscosity is $\eta_i = 1$~mPa\,s. The oil viscosity is measured with a homemade capillary viscometer at $\eta_o = 120$~mPa\,s. Thus, the viscosity ratio is $\lambda = \eta_i/\eta_o= 0.008$. The interfacial tension between water and oil, measured using the pendant drop technique~\cite{DelRio1997}, is $\gamma = 48$~mN/m. Fluids are injected at constant flow rates using syringe pumps (Legato 180, KdScientific). The microfluidic setup is mounted in an upright microscope (Eclipse 50i, Nikon) and imaged with a CCD camera (Marlin F-033B, Allied) at 30 frames per second. The shape of the drops is determined by image analysis with custom made Matlab software.

\subsection{\label{subsec:PIV}Two-dimensional hyperbolic flow}

The channel geometry is designed to produce a depth-averaged hyperbolic flow of the form
\begin{equation}
\label{eq:Hyperbolic}
u = -Gx, \quad v = Gy.
\end{equation}
in the test section in the absence of drops. In Eq.~(\ref{eq:Hyperbolic}), $u$ and $v$ are the depth-averaged velocity components in the $x$ and $y$ directions defined in Fig.~\ref{fig:Channel}(b). $G$ is the strain rate of the flow and is constant for given $Q_{\rm tot}$ and channel geometry. In particular, we expect $G$ to depend linearly on $Q_{\rm tot}$. For a given total flow rate $Q_{\rm tot}$, the mean oil velocity flowing into the test section from each side can be computed as $U = Q_{\rm tot}/hw$ and from it a characteristic strain rate can be defined as $G = U/w = Q_{\rm tot}/hw^2$. However, the presence of the channel walls alters this flow and Eq.~(\ref{eq:Hyperbolic}) is expected to stand only in a region near the center of the test section.

\begin{figure}[h]
\includegraphics[width=8cm]{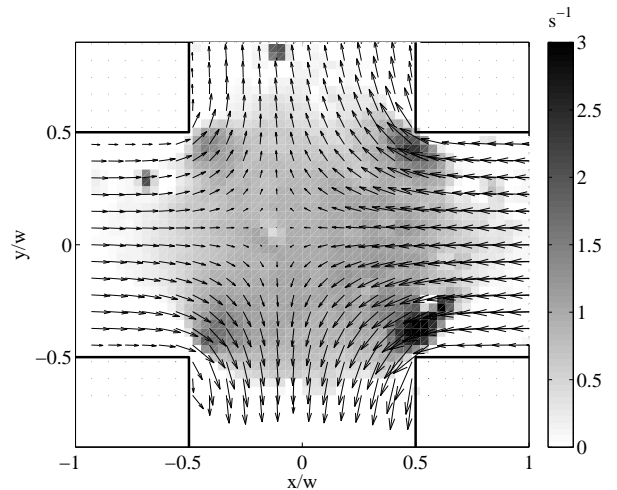}
\caption{\label{fig:PIV}Velocity field (arrows) and strain rate (gray scale) in the test region. Solid lines are drawn at the position of the channel walls.}
\end{figure}

To characterize the flow in the test section in the absence of drops, $1 \ \mu$m-diameter yellow-green fluorescent beads (FluoSpheres, Invitrogen) suspended in water are injected in the channel. A typical depth-averaged velocity field in the test section is measured with particle image velocimetry (PIV) in channel II and shown by the vector field in Fig.~\ref{fig:PIV}. In this case $Q_{\rm tot} = 0.4 \ \mu$L/min and hence $U = 0.29$~mm/s and $G = 0.7$~s$^{-1}$. The velocity field resembles the flow defined in Eq.~(\ref{eq:Hyperbolic}), with a stagnation point near the center of the test section.

The strain rate is computed from the PIV measurement as $-1/2(\partial u/\partial x -\partial v/\partial y)$ and is plotted in gray scale in Fig.~\ref{fig:PIV}. Excluding the four inner corners, the strain rate is nearly constant everywhere in the test section with a value of around $1$~s$^{-1}$. This value is consistent with the expected value $G = 0.7$~s$^{-1}$.

The vorticity, computed as $(\partial u/\partial y - \partial v/\partial x)$, remains negligible within the whole channel except the four inner corners of the test section (not shown). All these observations support the assumption of a hyperbolic flow [Eq.~(\ref{eq:Hyperbolic})] in the test section and negligible effects of the walls except at the inner corners of the test section. Henceforth, for each experiment the strain rate is computed from the total flow rate as $G = Q_{\rm tot}/hw^2$.

%%%%%%%%%%%%%%%%
% Results
%%%%%%%%%%%%%%%%

\section{\label{sec:Results}Results}

For given flow conditions (ie. drop size and strain rate) several drops are imaged as they flow through the microchannel. When drops enter the test section they deform, adopting an elongated shape. No drop breakup was observed in the range of drop sizes and strain rates used in these experiments. Instead, drops flow into the test section, decelerate while approaching the stagnation point and deform in the extensional flow. Eventually, drops leave the stagnation point, whether due to the arrival of the next drop or due to random flow fluctuations and leave the channel through one of the exits.

\begin{figure}[h]
\includegraphics[width=6cm]{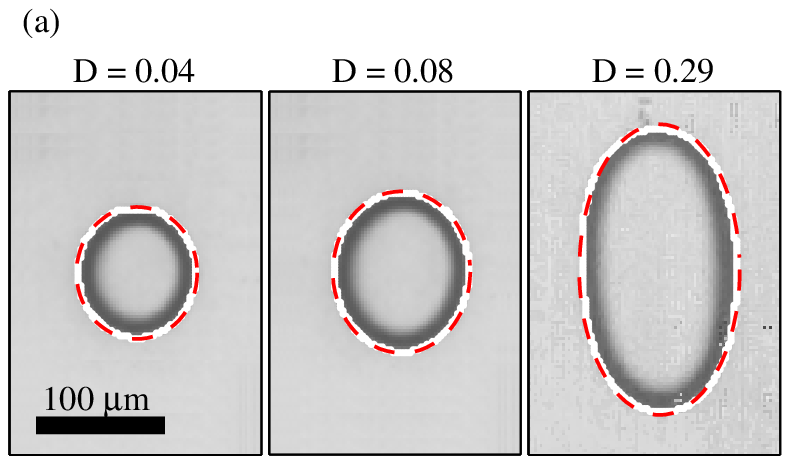}
\includegraphics[width=8cm]{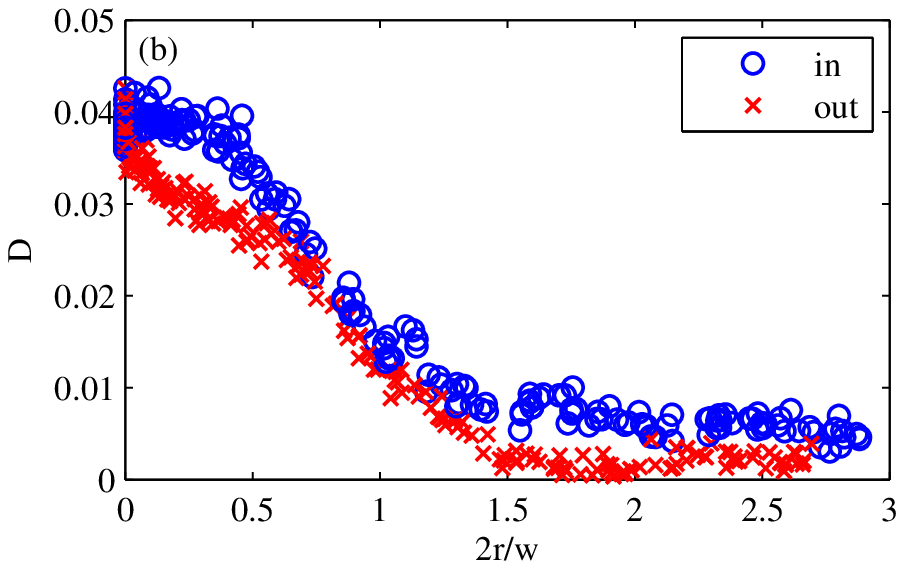}
\caption{\label{fig:D-evolution}(Color online) (a) Images of deformed drops. The white thick curve shows the digitally obtained contour of the drop. The red thin dashed curve corresponds to a fitted ellipse. (b) Deformation of a drop as it travels through the microchannel. $D$ is plotted against the normalized distance between the drop and the test section centers when the drop flows into (blue circles) and out of (red crosses) the test section. This case corresponds to data from channel I, with $R_o = 58 \ \mu$m and $G = 9.1$~s$^{-1}$.}
\end{figure}

The deformation of the drops is measured as a function of their position to the center of the test section $r = \sqrt{x^2 + y^2}$. For that, the contour of the drops is digitized and fitted with ellipses, as shown in Fig.~\ref{fig:D-evolution}(a) for three different cases. It can be seen that the elliptic shape provides an excellent fit for the drop contour that only slightly degrades for the most elongated drops. This is consistent with recent studies of drop shape relaxation~\cite{Brun2013}. The deformation of the drops is quantified from the major and minor axes of the fitted ellipses, $a$ and $b$ respectively, as
\begin{equation}
\label{eq:D-parameter}
D = \frac{a - b}{a + b}.
\end{equation}
Figure~\ref{fig:D-evolution}(b) shows a typical evolution curve of $D$ as a function of $2r/w$. Blue circles show the evolution of $D$ as drops flow into the test section while red crosses are used for drops leaving it. Drops start to deform when still outside the test section at $2r/w \approx 1.5$. The deformation increases as drops approach the stagnation point and saturates at a maximum value $D_{\rm max}$. As drops travel out of the test section, they rapidly relax back into a circular shape. Note that the curves of deformation when drops flow into and out of the test section almost coincide but differ slightly near $r = 0$. This is due to the different effect of the external flow in drop deformation into and out of the hyperbolic flow. Drops flowing into the test section experience a diverging external flow, while drops flowing out of the test section are subjected to a convergent flow. This  exemplifies the temporal symmetry breaking introduced by the presence of fluid interfaces: the flow without drops has temporal symmetry but it is no longer reversible in the presence of drops.

\begin{figure}[h]
\includegraphics[width=8cm]{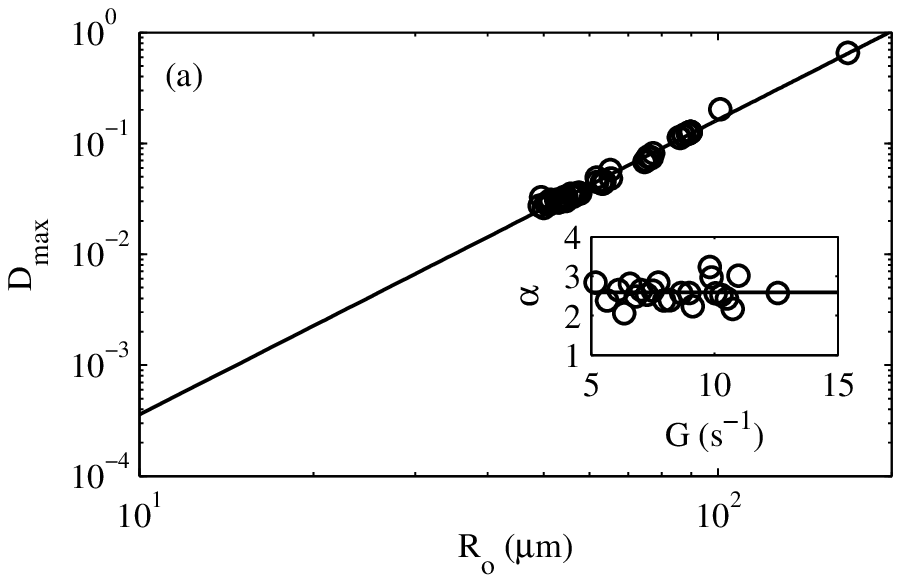}
\includegraphics[width=8cm]{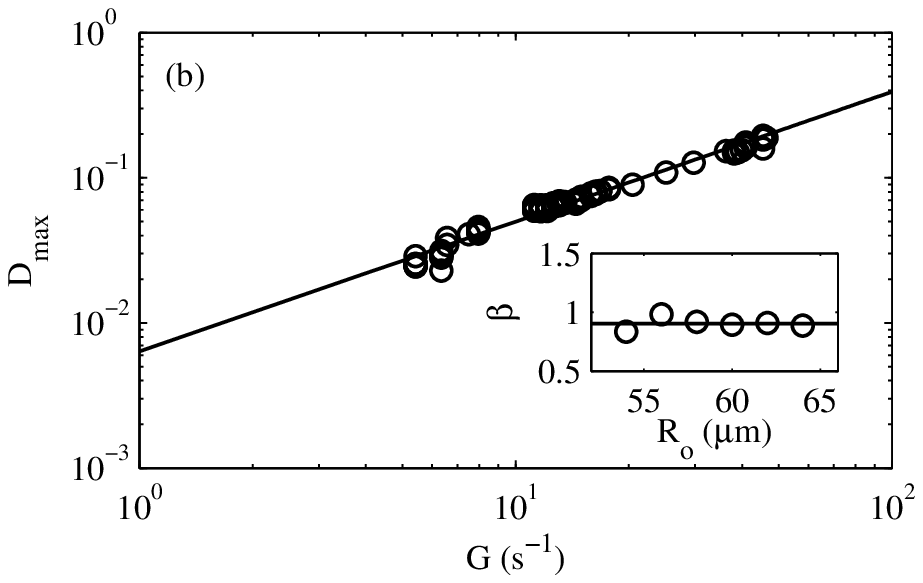}
\caption{\label{fig:DmaxScaling}(a) Maximum deformation of drops $D_{\rm max}$ as a function of drop size for constant strain rate $G = 12.6$~s$^{-1}$ in channel II. (b) $D_{\rm max}$ as a function of strain rate for fixed drop size $R_o = 60 \ \mu$m in channel I. In (a) and (b) the symbols correspond to experimental data and the solid lines correspond to power law fits. Insets: (a) Dependence of the fit coefficient $\alpha$ with strain rate $G$. (b) Dependence of fit coefficient $\beta$ with drop size $R_o$. The symbols represent fit data and the solid line shows the average.}
\end{figure}

The maximum drop deformation $D_{\rm max}$ is measured for each experimental condition. A typical curve of maximum deformation $D_{\rm max}$ as a function of $R_o$ for fixed $G$ is shown in Fig.~\ref{fig:DmaxScaling}(a). Similarly, a typical curve of $D_{\rm max}$ as a function of $G$ for fixed $R_o$ is shown in Fig.~\ref{fig:DmaxScaling}(b). It is observed that $D_{\rm max}$ increases if whether the drop size $R_o$ or the strain rate $G$ increase. The linear trend in log-log scale in both cases suggest that $D_{\rm max} \sim R_o^\alpha G^\beta$. It is important to note that geometrical constraints prevent us from achieving one full decade in the range of $R_o$ in Fig.~\ref{fig:DmaxScaling}(a).

In order to quantify the dependence of $D_{\rm max}$ on the drop size and strain rate, power laws are fit to the data. For each strain rate, a curve of the form $D_{\rm max} = C_1 R_o^\alpha$ is adjusted, with $C_1$ and $\alpha$ two fitting coefficients. The coefficients $\alpha$ for different strain rates $G$ are shown in the inset of Fig.~\ref{fig:DmaxScaling}(a). It is found that $\alpha$ has little dispersion around a mean value ($\pm$ standard deviation) $\overline{\alpha} = 2.59 \pm 0.28$.

Similarly, power law curves of the form $D_{\rm max} = C_2 G^\beta$ are fit for each drop size, with $C_2$ and $\beta$ as fitting coefficients. The coefficients $\beta$ for different drop sizes $R_o$ are shown in the inset of Fig.~\ref{fig:DmaxScaling}(b). Again, $\beta$ remains approximately constant for the different drop sizes. The average $\pm$ standard deviation is $\overline{\beta} = 0.91 \pm 0.05$.

%%%%%%%%%%%%%%%%
% Theory
%%%%%%%%%%%%%%%%

\section{\label{sec:Theory}Theory}

The deformation of drops in pure straining flows, simple shear flows and flows with varying rates of strain and vorticity has been largely studied since the seminal work of Taylor in the four-roll mill experiments~\cite{Taylor1934, Rallison1984, Stone1994}. Most studies, however, deal with three-dimensional drops in two-dimensional or axisymmetric hyperbolic flows. In 1968, Richardson considered two-dimensional bubbles~\cite{Richardson1968} and in 1973 Buckmaster and Flaherty expanded the theory to viscous drops~\cite{Buckmaster1973}. They take advantage of the use of analytical functions and conformal mappings to describe the flow. Here, we employ a similar approach but with the difference that our drops are not truly two-dimensional but are squeezed in a shallow geometry.

The theoretical model is based on the Darcy approximation of the Stokes equations for the depth-averaged velocity field $\vec u = u(x,y) \hat x + v(x,y) \hat y$ in a shallow geometry of height~$h$~\cite{Batchelor}:
\begin{equation}
\label{eq:HeleShaw}
\vec u = -\frac{h^2}{12 \eta} \nabla p.
\end{equation}
For an incompressible flow ($\nabla \cdot \vec u = 0$), these equations allow convenient solutions in terms of a complex potential $w(z) = \phi + i \psi$, in which $z = x + iy$ is the complex variable that defines the position in the $(x,y)$ plane, $\phi(x,y) = -h^2/(12\eta) p(x,y)$ is the velocity potential and $\psi(x,y)$ is the two-dimensional stream function. The fluid velocity is related to $w(z)$ as $w'(z) = u - iv$.

\begin{figure}[h]
\includegraphics[width=8cm]{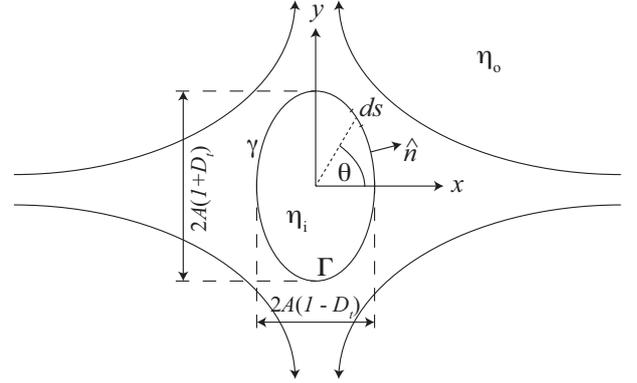}
\caption{\label{fig:TheoreticalDrop}A quasi two-dimensional drop subjected to a hyperbolic flow.}
\end{figure}

Now, consider the situation depicted in Fig.~\ref{fig:TheoreticalDrop}. A squeezed drop of viscosity $\eta_i$ is centered at the origin $x = y = 0$ immersed in a carrier fluid of viscosity $\eta_o$. The interfacial tension between both fluids is $\gamma$. A quasi two-dimensional hyperbolic flow of constant strain rate $G$ and zero vorticity, represented by Eq.~(\ref{eq:Hyperbolic}), is imposed on the unbounded outer fluid. Since the presence of the drop alters the external flow field, Eq.~(\ref{eq:Hyperbolic}) is only valid far from the drop and therefore:
\begin{equation}
\label{eq:Infinity}
w(z) \sim -\frac{Gz^2}{2} + O(z^{-2}) \quad \textrm{when } z \rightarrow \infty.
\end{equation}
Logarithmic terms are omitted because the pressure has to remain finite and odd powers of $z^{-1}$ do not to appear due to the symmetry of the problem.

The problem is to find the shape of the drop $\Gamma$ and the complex potentials inside and outside of the drop ($w_i(z)$ and $w_o(z)$ respectively) that satisfy the continuity of the depth-averaged velocity field and normal forces in $\Gamma$ and Eq.~(\ref{eq:Infinity}) at infinity. At this point we limit ourselves to the steady-state solution, in which the stream function is equal to a constant in $\Gamma$. Since this constant can be absorbed in the imaginary part of the complex potentials, the steady-state condition is:
\begin{equation}
\label{eq:Steady}
w(z) = \overline{w(z)} \quad \textrm{in } \Gamma,
\end{equation}
where the bar denotes complex conjugate. The continuity of velocity is written as:
\begin{equation}
\label{eq:Velocity}
w_o'(z) = w_i'(z) \quad \textrm{ for } z \textrm{ in } \Gamma,
\end{equation}
and the jump in pressure due to the interface curvature is expressed, with the aid of Eq.~(\ref{eq:Steady}), as~\cite{Richardson1968, Park1984}:
\begin{equation}
\label{eq:Tangency}
\frac{12i}{h^2}\left[\eta w(z) \right]_i^o dz = \gamma \left( d \left( \frac{dz}{ds} \right) + \frac{2}{h} i dz \right), \quad z \textrm{ in } \Gamma.
\end{equation}
The brackets mean the difference of the expression outside and inside the drop and the arc length $s$ is defined in Fig.~\ref{fig:TheoreticalDrop}. The two terms in the right hand side of Eq.~(\ref{eq:Tangency}) account for the effects of the curvature of the interface, both in the $(x, y)$ plane and in the perpendicular direction, respectively. In particular, the curvature in the direction normal to the plane, which we assume to remain constant equal to $2/h$, gives rise to a constant over-pressure inside the drop that can be absorbed in the real part of $w_i$. Therefore, in the following, we will neglect the second term in Eq.~(\ref{eq:Tangency}) and keep in mind that $w_i$ contains only a portion of the pressure inside the drop.

Equation~(\ref{eq:Tangency}) can be integrated by defining $\xi_{i,o}'(z) = w_{i,o}(z)$. The problem is simplified by assuming $\eta_i = \eta_o = \eta$. With this, the pressure jump at the drop interface can be written as
\begin{equation}
\frac{12i\eta}{h^2} \left( \xi_i(z) - \xi_o(z) \right) = -\gamma \frac{dz}{ds} \quad z \textrm{ in } \Gamma.
\end{equation}
Therefore, by the Sokhotski-Plemelj formula, the velocity potential for any $z$ can be computed as~\cite{Buckmaster1973}:
\begin{equation}
\label{eq:Plemelj}
w(z) = \frac{h^2 \gamma}{24 \pi \eta} \int_\Gamma \frac{dt/ds}{(t-z)^2} dt - \frac{Gz^2}{2}% + O(z^{-1}).
\end{equation}

In conclusion, we have the problem of finding the shape of the drop $\Gamma$ such that the complex potential of Eq.~(\ref{eq:Plemelj}) has continuous derivative [Eq.~(\ref{eq:Velocity})] and is real [Eq~(\ref{eq:Steady})] in $\Gamma$. The mathematical difficulty of this problem is high. Instead we solve the problem in an approximate way. For that, we suppose that the drop assumes the shape of an ellipse of semi-axes $A(1+D_t)$ and $A(1-D_t)$, with $0 < D_t < 1$, as shown in Fig.~\ref{fig:TheoreticalDrop}. Note that the elliptic drop shape is introduced by means of a conformal map and therefore other drop shapes could be considered by means of appropriately defined conformal maps. 

Proceeding as in Ref.~\cite{Buckmaster1973} we find an integral equation for $D_t$:
\begin{equation}
\label{eq:Integral}
\frac{-1}{3\pi A} \int_\Gamma \frac{dt}{ds} dt = \frac{4 \eta G A^3}{\gamma h^2} (1-D_t^2) = \frac{4 \eta G R_o^3}{\gamma h^2\sqrt{1-D_t^2}}.
\end{equation}
In the last expression, the conservation of drop volume was imposed in the form $\pi R_o^2 = \pi A^2(1-D_t^2)$, where $R_o$ is the radius of the non-deformed drop. The integral in the left hand side of Eq.~(\ref{eq:Integral}) is proportional to $A$ and therefore this is an equation for $D_t$ with one non-dimensional group:
\begin{equation}\label{eq:P-Parameter}
P = \frac{4 \eta G R_o^3}{\gamma h^2} = Ca \, \delta^2,
\end{equation}
that involves the capillary number $Ca = \eta G R_o/\gamma$ and the confinement parameter $\delta$. Finally, the equation for $D_t$ is found by expressing the integral of Eq.~(\ref{eq:Integral}) at the surface of the ellipse:
\begin{equation}
\label{eq:DFundamental}
\frac{1}{3\pi} \int_0^{2\pi} \frac{2D_t + (1+D_t^2) \cos(2\theta)}{\sqrt{1+D_t^2 + 2D_t \cos(2\theta)}} d\theta = \frac{P}{\sqrt{1-D_t^2}}.
\end{equation}

\begin{figure}[h]
\centering
\includegraphics[width=8cm]{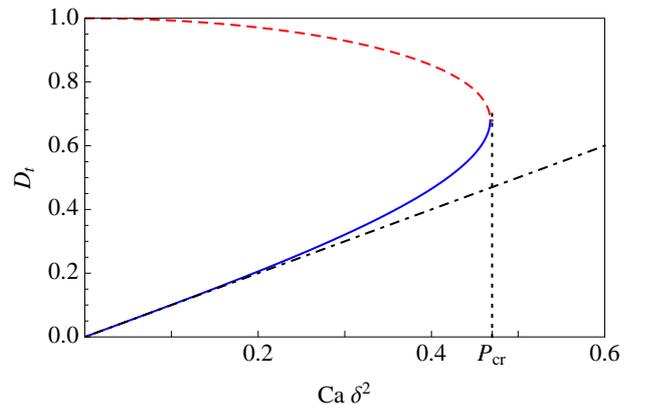}
\caption{\label{fig:TheoD}(Color online) Solution of Eq.~(\ref{eq:DFundamental}). The lower branch (solid blue line) corresponds to the physical solution, while the upper branch (dashed red line) is unphysical. The linear relationship $D_t = P$ is plotted in black dash-dotted line for comparison. The vertical black dotted line shows the position of $P_{\rm cr}$.}
\end{figure}

Equation~(\ref{eq:DFundamental}) is solved numerically for $D_t$ as a function of $P$ and the result is plotted in Fig.~\ref{fig:TheoD}. For values of $P < P_{\rm cr} = 0.47$ two branch solutions for $D_t$ exist. The lower branch is well approximated by $D_t = P$ for small $D_t$. For larger $P$, it deviates from the linear trend, first slightly and then more rapidly. In the second branch, $D_t$ decreases with increasing $P$ from $D_t(P = 0) = 1$, first slightly and then more rapidly. Since the limit of zero capillary number, hence $P = 0$, corresponds to zero strain rate, it seems that the physical solution corresponds to the lower branch.

Both branchs meet at $D_t(P_{\rm cr}) = 0.68$, beyond which there is no solution for $D_t$. Nonexistence of solution for the deformation parameter has been related to bursting of the drops~\cite{Richardson1968, Buckmaster1973} and therefore the theoretical model predicts drop breakup for $P > P_{\rm cr}$ for viscosity ratio of unity.

%%%%%%%%%%%%%%%%
% Discussion
%%%%%%%%%%%%%%%%

\section{\label{sec:Discussion}Discussion}

The model presented above, based on the Darcy approximation for shallow geometries, predicts the deformation of squeezed drops in a hyperbolic flow for unit viscosity ratio. The model states that the steady-state deformation of drops $D_t$ depends on a non-dimensional parameter $P$ given by Eq.~(\ref{eq:P-Parameter}) and that for small deformations, $D_t = P$. In other words, the model predicts that, for small deformations, $D_t \sim G R_o^3$. Our experimental results for the maximum drop deformation, on the other hand, show that $D_{\rm max} \sim R_o^{2.59} G^{0.91}$. This experimental scaling is in good agreement with the theoretical model in the small deformations limit. It is important to note that the model predicts the steady shape of the drop, while our experiments are performed on flowing drops. However, as shown in Fig.~\ref{fig:D-evolution}, the deformation of drops rapidly saturates at the maximum value $D_{\rm max}$, which represents, therefore, the steady-state drop deformation.

The parameter $P$ can be understood by considering the forces involved in the deformation of drops. In shallow geometries it is not the viscous stress but the pressure of the external fluid which has a larger influence on drop deformation. As evidenced by Eq.~(\ref{eq:HeleShaw}) the pressure in the outer fluid scales as $\eta_o G (R_o/h)^2$ while viscous stresses scale only as $\eta_o G$. On the other hand, the interfacial forces that resist the deformation are proportional to the curvature of the drop interface and it has been argued that only the curvature in the plane affects the dynamics of the drop. Thus, the interfacial forces per unit area scale as $\gamma/R_o$. The competition between these forces yields the parameter $P = Ca \, \delta^2$.

\begin{figure}[t]
\includegraphics[width=8cm]{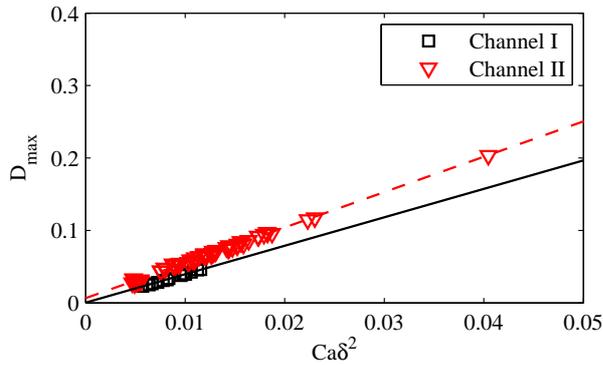}
\caption{\label{fig:Dmax_P}(Color online) $D_{\rm max}$ as a function of $P = Ca \, \delta^2$ for different drop sizes and strain rates from data taken in the two channels used on the study (see Section~\ref{subsec:Setup}). The symbols correspond to experimental data, the black solid line and the red dashed line correspond to linear fits of the data from channel I ($D_{\rm max} = 3.9 P$) and channel II ($D_{\rm max} = 5 P$) respectively.}
\end{figure}

In order to further test the predictions of the theory with the experimental data, the measurements of $D_{\rm max}$ for different strain rates and drop sizes are plotted as a function of $P$ in Fig.~\ref{fig:Dmax_P}. The data for each channel collapse in single linear curves with slightly different slopes, $3.9$ for channel I and $5$ for channel II. The slopes are larger than the prediction of the theory, which predicts a slope of unity. One possible reason for this discrepancy is the different viscosity ratio used in the experiments, $\lambda = 0.008$, meaning that less viscous drops are more deformable. Indeed, it would be interesting to explore the effect of viscosity ratio on drop deformation. Another possible explanation is the influence of three-dimensional effects due to the finite degree of confinement. The analysis of force competition presented above is justified under the assumption of high confinement $\delta \gg 1$, since in that case the pressure terms are much larger than viscous stresses and three-dimensional effects are negligible. Therefore, the difference in the slopes of the linear fits in Fig.~\ref{fig:Dmax_P} could be due to neglecting small but not insignificant viscous stresses at the drop interface or due to three-dimensional effects since we do not work in the limit $\delta \gg 1$. The channel depth, however, alters the slope of the linear relationship between the deformation and the parameter $P$ but not the scaling $D_{\rm max} \propto P$ in the limit of small deformations.

Two-dimensional depth-averaged models such as the one presented here cannot describe the complex three-dimensional effects that occur at the interface between immiscible fluids~\cite{Sarrazin2006, Jakiela2012}. However, as suggested by two-dimensional numerical simulations~\cite{Hoang2013} and shown by this work, they are able to describe the flow of biphasic flows with reasonable accuracy in many situations. We believe that this kind of models can be of valuable use in the description of droplet microfluidics when confinement is important.

%%%%%%%%%%%%%%%%
% Acknowledgments
%%%%%%%%%%%%%%%%

\acknowledgments{This work was funded by CONICYT through FONDECYT Iniciaci\'on grant No. 11100204 and Anillo de Investigaci\'on en Ciencia y Tecnolog\'ia grant No. ACT 127.}

%\bibliographystyle{apsrev}
%\bibliography{../../../Bibliografia/Bibliografia}

\end{document}